\documentclass[twocolumn,showpacs,preprintnumbers,amsmath,superscriptaddress,amssymb,pra,aps]{revtex4-1}

\usepackage{color}
\usepackage{graphicx}
\usepackage{dcolumn}
\usepackage{bm}
\newcommand{\mr}{\mathrm}

\begin{document}


\title{Effect of disorder close to the superfluid transition in a two-dimensional Bose gas}

\author{B. Allard}
\affiliation{Laboratoire Charles Fabry, Institut d'Optique, CNRS, Univ Paris-Sud, 2, Avenue Augustin Fresnel, 91127 PALAISEAU
CEDEX}
\author{T. Plisson}
\affiliation{Laboratoire Charles Fabry, Institut d'Optique, CNRS, Univ Paris-Sud, 2, Avenue Augustin Fresnel, 91127 PALAISEAU
CEDEX}
\author{M. Holzmann}
\affiliation{LPTMC, UMR 7600 of CNRS, Universit\'e P. et M. Curie, 75752 Paris, France}
\affiliation{Universit\'e Grenoble 1/CNRS, LPMMC, UMR 5493, B.P. 166, 38042 Grenoble, France}
\author{G. Salomon}
\affiliation{Laboratoire Charles Fabry, Institut d'Optique, CNRS, Univ Paris-Sud, 2, Avenue Augustin Fresnel, 91127 PALAISEAU
CEDEX}
\author{A. Aspect}
\affiliation{Laboratoire Charles Fabry, Institut d'Optique, CNRS, Univ Paris-Sud, 2, Avenue Augustin Fresnel, 91127 PALAISEAU
CEDEX}
\author{P. Bouyer}
\affiliation{Laboratoire Charles Fabry, Institut d'Optique, CNRS, Univ Paris-Sud, 2, Avenue Augustin Fresnel, 91127 PALAISEAU
CEDEX}
\affiliation{LP2N, Univ Bordeaux 1, IOGS, CNRS, 351 cours de la Lib\'eration, 33405 Talence, France}
\author{T. Bourdel}
\email[Corresponding author: ]{thomas.bourdel@institutoptique.fr}
\affiliation{Laboratoire Charles Fabry, Institut d'Optique, CNRS, Univ Paris-Sud, 2, Avenue Augustin Fresnel, 91127 PALAISEAU
CEDEX}

\date{\today}

\begin{abstract}
We experimentally study the effect of disorder on trapped quasi two-dimensional (2D) $^{87}$Rb clouds in the vicinity of the Berezinskii-Kosterlitz-Thouless (BKT) phase transition. The disorder correlation length is of the order of the Bose gas characteristic length scales (thermal de Broglie wavelength, healing length) and disorder thus modifies the physics at a microscopic level. We analyze the coherence properties of the cloud through measurements of the momentum distributions, for two disorder strengths, as a function of its degeneracy. For moderate disorder, the emergence of coherence remains steep but is shifted to a lower entropy. In contrast, for strong disorder, the growth of coherence is hindered. Our study is an experimental realization of the dirty boson problem in a well controlled atomic system suitable for quantitative analysis.
\end{abstract}

\pacs{74.62.En, 
05.60.Gg, 
67.85.Jk, 
05.10.Ln 
}

\maketitle
 
 Together with band structure and interactions, disorder is a key ingredient for the understanding of transport in condensed matter physics \cite{Ashcroft76}. At low temperature, it affects the conductivity of a metal and it even induces phase transitions to insulating states \cite{Imada98}. A striking example is Anderson localization \cite{Anderson58}, which has recently been observed in 3D ultracold gases \cite{Kondov11}.

Disorder is especially relevant in 2D systems, such as Si-MOSFET \cite{Kravchenko04}, or thin metal films \cite{Goldman98}, in which quantum phase transitions to insulating phases have been observed. Moreover, in high-Tc superconductors, doping intrinsically introduces inhomogeneities in the CuO-planes \cite{Pan01}. Understanding the complex interplay between disorder and interactions in these systems remains a major challenge. 

Whereas the above mentioned electronic systems are fermionic, superconductivity originates from the bosonic nature of Cooper pairs. As long as disorder does not break the Cooper pairs, the problem is reduced to a study of dirty bosons \cite{Fisher90, Bollinger11}. It has mainly been studied numerically in the framework of the disordered 2D Bose-Hubbard model. Disorder can both favor or disfavor superfluidity \cite{Krauth91}, and the occurrence of a Bose glass, an insulating, gapless, compressible phase has been predicted \cite{Fisher90}.
      
In the context of ultra-cold atoms, the properties of disordered $trapped$ Bose gases have been studied both in 1D \cite{Fallani07, Chen08, Gadway11} and 3D \cite{White09}. In 2D Bose gases in the absence of disorder, the Berezinskii-Kosterlitz-Thouless (BKT) superfluid phase transition \cite{Berezinskii72} has been experimentally studied through the modification of the gas coherence properties associated with the pairing of thermal vortices \cite{Hadzibabic06, superfluid}. In a continuous system, the effect of disorder on the BKT superfluid transition is expected to depend on the correlation length of the disorder $\sigma$, which has to be compared to the characteristic length scales of the cloud such as the thermal de Broglie wavelength $\lambda_\mr{dB}$ and the healing length $\xi$, $i.e.$ the vortex core size \cite{Pilati09}. For slowly varying disorder ($\sigma \gg \xi, \lambda_\mr{dB}$), the physics can be locally described by the homogeneous BKT transition, and disorder causes a percolation transition of superfluid islands. In contrast, for a microscopically correlated disorder ($\sigma \lesssim \xi, \lambda_\mr{dB}$), tunneling is possible and the very nature of the phase transition is affected. 

A recent Monte-Carlo study of the homogeneous 3D Bose gas in the presence of speckle disorder \cite{Pilati09} has shown that, depending on its correlation length $\sigma$,  the disorder can either reduce the critical temperature (due to quantum localization) or increase it because of the reduction of the available volume (see Fig.\,14 in \cite{Pilati09}). However there have been so far no theoretical prediction for the effect of disorder on the BKT superfluid transition in a continuous 2D system. As in 3D, Anderson localization \cite{Anderson58} and percolation phenomena are likely to affect the superfluid transition, but reducing dimensionality should enhance their effects \cite{Abrahams79, Pilati09}. In addition, new phenomena affecting specifically the BKT transition such as enhanced phase fluctuations or vortex pinning \cite{Blatter94} may play an important role.
 
In this letter, we present an experimental study of the effect of microscopically correlated disorder on the coherence properties of a 2D ultracold atomic gas near the BKT superfluid transition. As in \cite{Plisson11}, the coherence properties are probed by the study of the momentum distribution, which is the Fourier transform of the first order correlation function $g_{1}$ \cite{Gerbier03}. We observe that an adiabatic ramping up of the disorder results in a suppression of the low momentum peak, $i.e.$ a decrease of coherence. In particular, for a moderate disorder strength of 0.4 times the temperature, we measure a small shift of the emergence of coherence towards low entropy. For stronger disorder strength of the order of the temperature, the growth of coherence is significantly hindered both as a function of entropy and temperature.

Our experiment starts with a quantum degenerate 2D Bose gas in a trap obtained from a combination of a blue detuned TEM$_{01}$ beam, which confines the gas in a horizontal plane, and a red detuned Gaussian beam for the in-plane confinement \cite{Plisson11}. The trap oscillation frequencies are  $\omega_{x}/2\pi = 8\,\rm{Hz}$, $\omega_{y}/2\pi = 15\,\rm{Hz}$, $\omega_{z}/2\pi = 1.5\,\rm{kHz}$. The atom number $N$ is varied between $2\times 10^{4}$ and $6\times 10^{4}$ in order to change the degeneracy of the gas across the BKT transition. The temperature, measured from a fit to the wings of the momentum distribution using a Hartree-Fock mean-field model \cite{Plisson11}, remains constant at 64.5$\pm 2.0\,$nK. At this temperature, $\sim 70 \%$ of the atoms are in the ground state of the vertical harmonic oscillator. The dimensionless 2D interaction strength is $\tilde{g}=\sqrt{8\pi}a_{\rm{s}}/a_{\rm{z}}=0.096$, where $a_{\rm{s}}=5.3\,\rm{nm}$ is the 3D scattering length, $a_{\rm{z}}=\sqrt{\hbar/m\omega_{z}} \approx 0.28\,\mu$m the vertical harmonic oscillator characteristic length, $m$ the atom mass, and $\hbar$ the reduced Planck constant. 

The disorder potential is a speckle pattern produced by a 532\,nm laser beam, which passes through a diffusive plate and is focused on the atoms. The repulsive disorder potential is characterized by its mean value $\overline{V}$ (equal to its standard deviation) and its correlation lengths, inversely proportional to the numerical aperture of the optical system \cite{Clement06}. Given the intensity of the beam and its transverse waist radius of 1\,mm, the maximum value of $\overline{V}$ felt by the atoms is $\overline{V}_{\rm{max}} =k_\textrm{B}\times 60(10)\,\rm{nK}$. As the beam is tilted by $30^\circ $, the in-plane disorder is effectively anisotropic \cite{Robert10}. The correlation lengths of the disorder are such that $\sigma_{x}/2 = \sigma_{y} = 0.5\, \mu\rm{m}$ (half-width at $1/\sqrt{e}$). These correlation lengths are of the order of both the thermal de Broglie wavelength $\lambda_\mr{dB}=\sqrt{2 \pi \hbar^2/m k_\textrm{B}T} \approx 0.73\,\mu$m and the healing length (at the BKT transition) $\xi=\lambda_\mr{dB}/\sqrt{D_c \tilde{g}} \approx 0.82\,\mu$m, where $D_c \approx \log (380.3/\tilde{g}) \approx 8.3$ is the BKT critical phase space density \cite{Prokofev01}. However,  $\sigma_x$ and $\sigma_y$ are small compared to the Thomas-Fermi radii of the cloud at the BKT transition $l_x=\frac{\hbar \sqrt{2\tilde{g}D_c}}{m\omega_x \lambda_\mr{dB}}$=25\,$\mu$m, $l_y=(\omega_x/\omega_y) l_x=13\,\mu$m, a necessary condition for self-averaging measurements. In our experiment, we use a single realization of the speckle pattern.

In the experimental sequence, the disorder potential is slowly ramped up in 250\,ms after the preparation of the 2D gas. After a holding time of 250\,ms, all trapping potentials including the disorder are switched off and the atom cloud expands in 3D during a free fall of $83.5\,\rm{ms}$.  The column density of the gas along $z$ is then measured by fluorescence imaging from the top. As explained in \cite{Plisson11}, it reflects the in-trap momentum distribution in the  $x,y$ plane. Since, the momentum distributions appear to be cylindrically symmetric, we perform an azimuthal averaging \cite{Rath10} to obtain the momentum profiles $n(k)$ as a function of the wavenumber\,$k$.

\begin{figure}[t!]
\centering
\includegraphics[width=0.49\textwidth]{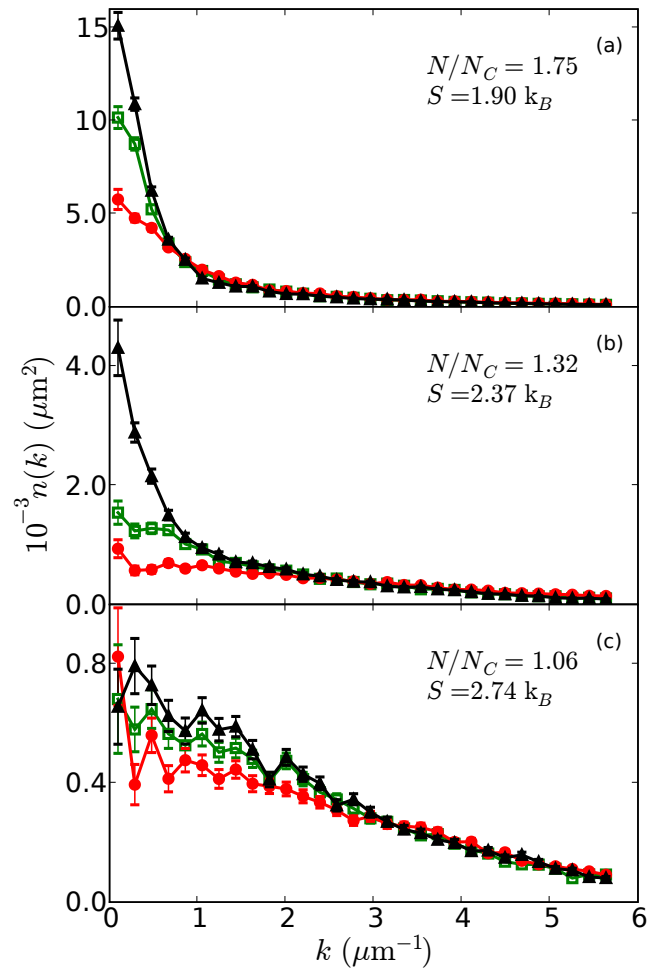}
\caption{\label{profils} (Color online) Azimuthally averaged momentum distribution profiles for  $N=5.6\times 10^{4}$ (a), $N=3.8\times 10^{4}$ (b) and $N=2.9 \times 10^{4}$ (c). In each case, we show the influence of the disorder\,: no disorder  $\overline{V}=0$ (black triangles), $0.4\,\overline{V}_{\rm{max}}$  (green open squares), $\overline{V}_{\rm{max}}$ (red circles).}
\end{figure}

We study the effect of disorder on the momentum distribution for different initial conditions, both above and below the BKT transition. A narrow momentum distribution is the signature of a large coherence length in the sample. We quantify the degeneracy of the non-disordered gas with the ratio $N/N_c$, where $N_c$ is the critical atom number of an ideal 3D Bose gas in our anisotropic trap. From our previous study of the 2D Bose gas in the absence of disorder \cite{Plisson11}, we know that the BKT phase transition happens at $N/N_c \approx 1.26$ for our parameters.  In Fig.\,\ref{profils}, we compare the momentum distribution without disorder to the results obtained after ramping up the disorder potential to $\overline{V}=0.4\,\overline{V}_{\rm{max}}$ and $\overline{V}=\overline{V}_{\rm{max}}$.  For $N/N_c =1.06$ (Fig.\,\ref{profils}c) and in the absence of disorder,  the gas is in the normal phase. In this case, the addition of the disorder has little effect, reducing slightly the low momentum population ($k<2\,\mu$m$^{-1}$). For $N/N_c =1.32$ (Fig.\,\ref{profils}b), the gas has just entered the superfluid phase in the absence of disorder and a low momentum peak is clearly present. In this case, the disorder has a strong effect. The population at very low momentum ($k<0.5\,\mu$m$^{-1}$) is strongly suppressed and the low momentum peak almost disappears. The profiles with disorder are then qualitatively similar to the one in the normal phase (Fig.\,\ref{profils}c). For $N/N_c =1.75$ (Fig.\,\ref{profils}a) and in the absence of disorder, the gas is deep in the superfluid phase with a large low momentum peak. In this case, the addition of the disorder leads to a reduction of the height of the peak but not to its disappearance. In all our data, adding disorder always results in a reduction of the coherence of the Bose gas. 

It should be noted that applying slowly the disorder preserves the entropy. When the disorder potential is ramped up to a mean value $\overline{V}_{\rm{max}}$ in 250\,ms and then down in 250\,ms, we find no heating and no atom loss compared to the non-disordered situation, within our experimental precision ($\pm 1\,$nK). This uncertainty comes from both the shot-to-shot fluctuation of the experiment and the accuracy of the fitting procedure. We thus observe that adding the disorder is a reversible process. In the following, we assume that consecutive pictures with and without disorder correspond to the same entropy (Fig.\,\ref{profils}). Presenting our data as a function of entropy is then a natural choice.

\begin{figure}[t!]
\centering
\includegraphics[width=0.49\textwidth]{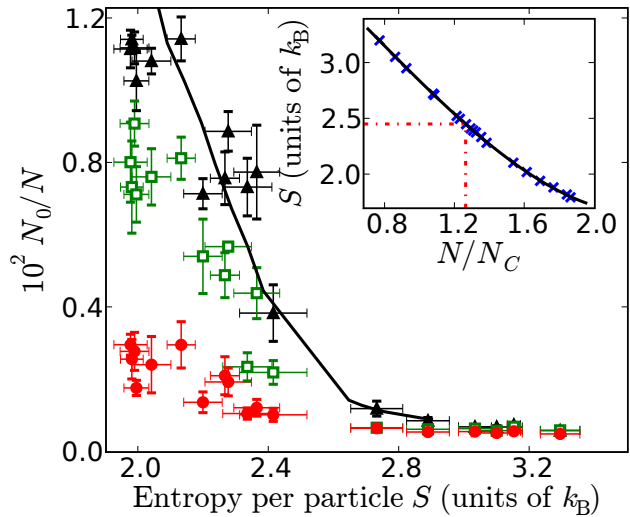}
\caption{\label{Scst} (Color online) Fraction of atoms ${N_{0}/N}$ in the central pixel of the momentum distribution as a function of the average entropy per particle $S$: non-disordered case $\overline{V}=0$ (black triangles), $\overline{V}=0.4\,\overline{V}_{\rm{max}}$ (green open squares), $\overline{V}=\overline{V}_{\rm{max}}$ (red circles). Each point results from the averaging of 5 experimental profiles and the error bars are statistical. The line corresponds to a Monte-Carlo simulation in the absence of disorder \cite{Plisson11}. Inset\,: entropy per particle measured by quantum Monte-Carlo simulations at 64.5nK (blue cross) and fitted by a $3^{rd}$ order polynomial (black line). The dashed lines indicate the BKT transition.}
\end{figure}

Even in the absence of disorder, we however do not have experimental access to the entropy. To find the correspondence between the ratio $N/N_c$, extracted from our measurements and entropy, we rely on quantum Monte-Carlo simulations of the non-disordered $in$-$situ$ distribution \cite{Holzmann08}, from which the entropy can be determined because of the scale invariance of the 2D Bose gas \cite{Yefsah11}. The calibration of the average entropy per particle $S$ as a function of $N/N_c$ for our experimental conditions is shown in the inset of Fig.\,\ref{Scst}. 

In order to analyze our result in a simple way, we would like to characterize the degree of coherence of the gas with a single number and not with the full momentum distribution. A natural quantity to consider is the coherence length, $i.e.$ the inverse of the width of the momentum distribution. However, in our case, the width of the momentum distribution saturates because of the limited resolution of our imaging system for highly coherent clouds \cite{Plisson11}. As an alternative, we choose to focus our analysis on the fraction of atoms $N_0/N$ in the central pixel of the momentum distribution ($k<0.2\,\mu$m$^{-1}$). It is also a well-defined model-independent quantity and it is related to the fraction of atoms that are coherent on a length scale larger than $\sim$5\,$\mu$m \cite{correspondence}. We plot  $N_0/N$ as a function of the entropy per particle $S$ (Fig.\,\ref{Scst}) and find that, at fixed entropy, the coherence of the gas is reduced in the presence of disorder. For $\overline{V}=0.4\,\overline{V}_{\rm{max}}$, the emergence of coherence is slightly shifted to a lower entropy per particle, by $0.2(1)\,k_\textrm{B}$, compared to the non-disordered case. For $\overline{V}=\overline{V}_{\rm{max}}$, we never reach a sufficiently low entropy to observe a large increase of coherence. 

Since the phase diagram of disordered systems is typically presented as a function of disorder and temperature \cite{Aleiner10}, we now complement our analysis of the coherence as a function of these quantities. This means that we have to determine the temperature from the experimental disordered profiles. In the absence of an exact theoretical model for 2D disordered gases, we use our non-disordered Hartree-Fock mean-field model \cite{Plisson11}, which we expect to be valid at large momenta. Experimentally, we fit to the wings of the distribution between a variable cut-off momentum $k_c$ and 12\,$\mu$m$^{-1}$. We find that, for $2.75\,\mu$m$^{-1} \leq k_c \leq 3.75\,\mu$m$^{-1}$, our signal to noise ratio is sufficient and the fitted temperature varies typically less than 1\,nK, indicating that our model is reasonably accurate in this range. We use $k_c=3.5\mu\rm{m}^{-1}$ in the following analysis. For $\overline{V}=\overline{V}_{\rm{max}}$ the temperature is found to increase on average by 5.5\,nK compared to the non-disordered case ($T=64.5\,$nK). 

\begin{figure}[t!]
\centering
\includegraphics[width=0.49\textwidth]{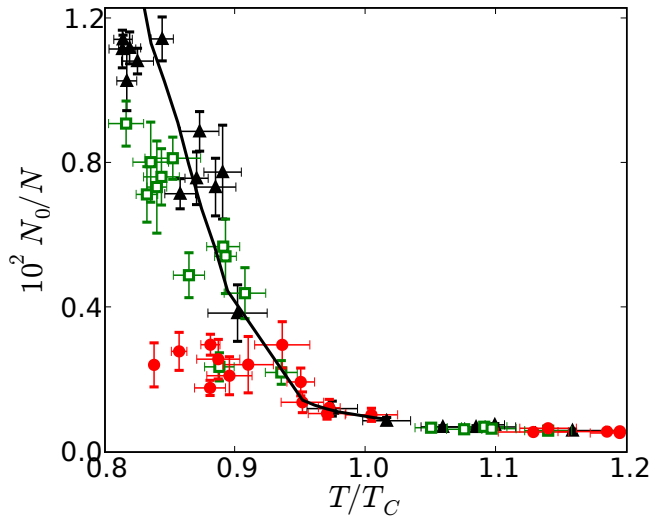}
\caption{\label{Tcoh}(Color online) Fraction of atoms ${N_{0}/N}$ in the central pixel of the momentum distribution as a function of the normalized estimated temperature for disorder strengths, $\overline{V}=0$ (black triangles), $\overline{V}=0.4\,\overline{V}_{\rm{max}}$ (green open squares), and  $\overline{V}=\overline{V}_{\rm{max}}$(red circles). $T_c$ is the critical temperature of an ideal 3D Bose gas in our anisotropic trap. Each point results from the averaging of 5 experimental profiles and the error bars are statistical. The line corresponds to a Monte-Carlo simulation in the absence of disorder \cite{Plisson11}.}
\end{figure}

Figure\,\ref{Tcoh} presents ${N_{0}/N}$ as a function of the temperature normalized to ${T_{c}}$, the critical temperature for an ideal 3D Bose gas in our anisotropic trap and the measured atom number. The results without disorder and with a disorder of amplitude  $0.4 \,\overline{V}_{\rm{max}}$ are  similar and no clear shift is visible. Within our accuracy, we can conclude that for this amount of disorder the coherence properties are weakly affected. Note that the classical percolation threshold $\sim0.52\,\bar{V}\approx 13\,$nK \cite{Weinrib82} is larger than the mean-field chemical potential at the phase transition in the absence of disorder $\mu_c=\tilde{g} \hbar^2 D_c /m {\lambda_\textrm{dB}}^2\approx k_\textrm{B} \times 8$\,nK. The observation of a weak effect of disorder is a signature that classical percolation is not relevant in the regime of a microscopically correlated disorder ($\sigma \sim \xi, \lambda_\mr{dB}$).

For $\overline{V}=\overline{V}_{\rm{max}}$,  the coherence increases much slower when $T/{T_{c}}$ decreases. The coherence properties of the gas are greatly modified. This finding contrasts with the 3D quantum Monte-Carlo calculation, which predicts a significant effect of the disorder only at larger disorder strength \cite{Pilati09}. We thus show the enhanced role of disorder in 2D as compared to 3D. At our lowest temperatures, ${N_{0}/N}$ does not reach the value of 0.0035 which corresponds to the superfluid transition in the absence of disorder \cite{Plisson11}. Although we do not directly measure the superfluid fraction \cite{superfluid}, since the superfluid transition is generally associated with the apparition of long range coherence, we can suspect that our system is not in a superfluid phase and that the critical temperature for superfluidity is shifted down by a significant amount. Actually, for our amount of disorder \cite{estimate}, the existence of a superfluid is not guaranteed even at zero temperature because of the disorder-driven quantum phase transition from a superfluid to an insulating Bose glass phase \cite{Aleiner10}.

In conclusion, we have shown that a microscopically correlated disorder ($\sigma \sim \xi, \lambda_\mr{dB}$) always reduces the coherence of a 2D Bose gas, both at constant temperature and entropy. For moderate disorder strength of 0.4\,$T$, the reduction is weak although we are able to measure a small shift of the emergence of coherence toward low entropy. A disorder strength of the order of gas temperature leads to a qualitative change of behavior with a suppressed coherence growth. We interpret the observed strong suppression of the coherence growth as a large shift of the superfluid transition. Theoretical studies in our experimental conditions would be of great interest and would permit to strengthen our analysis.
The mechanism of the disorder action can also be addressed both experimentally and theoretically. In the future, similar studies on strongly interacting fermions would add the possibility to elucidate the physics of disorder-induced breaking of bosonic pair.

We acknowledge F. Moron and A. Villing for technical assistance, J. Dalibard, T. Giamarchi, W. Krauth, and S. Piatecki for discussions. This research was supported by CNRS, Minist\`ere de l'Enseignement Sup\'erieur et de la Recherche, Direction G\'en\'erale de l'Armement, ANR-08-blan-0016-01, IXBLUE, RTRA: Triangle de la physique, EuroQuasar program of the ESF and EU, iSense, ERC senior grant Quantatop. LCFIO is member of IFRAF.

N.B.: In the last stage of the redaction of our paper, we have learned about a complementary work about disordered 2D Bose gases in the deep superfluid regime \cite{Beeler11}. It is focused on the different behaviors between quasi-condensate fraction and coherence.


\begin{thebibliography}{0}%
\makeatletter
\providecommand \@ifxundefined [1]{%
 \@ifx{#1\undefined}
}%
\providecommand \@ifnum [1]{%
 \ifnum #1\expandafter \@firstoftwo
 \else \expandafter \@secondoftwo
 \fi
}%
\providecommand \@ifx [1]{%
 \ifx #1\expandafter \@firstoftwo
 \else \expandafter \@secondoftwo
 \fi
}%
\providecommand \natexlab [1]{#1}%
\providecommand \enquote  [1]{``#1''}%
\providecommand \bibnamefont  [1]{#1}%
\providecommand \bibfnamefont [1]{#1}%
\providecommand \citenamefont [1]{#1}%
\providecommand \href@noop [0]{\@secondoftwo}%
\providecommand \href [0]{\begingroup \@sanitize@url \@href}%
\providecommand \@href[1]{\@@startlink{#1}\@@href}%
\providecommand \@@href[1]{\endgroup#1\@@endlink}%
\providecommand \@sanitize@url [0]{\catcode `\\12\catcode `\$12\catcode
  `\&12\catcode `\#12\catcode `\^12\catcode `\_12\catcode `\%12\relax}%
\providecommand \@@startlink[1]{}%
\providecommand \@@endlink[0]{}%
\providecommand \url  [0]{\begingroup\@sanitize@url \@url }%
\providecommand \@url [1]{\endgroup\@href {#1}{\urlprefix }}%
\providecommand \urlprefix  [0]{URL }%
\providecommand \Eprint [0]{\href }%
\providecommand \doibase [0]{http://dx.doi.org/}%
\providecommand \selectlanguage [0]{\@gobble}%
\providecommand \bibinfo  [0]{\@secondoftwo}%
\providecommand \bibfield  [0]{\@secondoftwo}%
\providecommand \translation [1]{[#1]}%
\providecommand \BibitemOpen [0]{}%
\providecommand \bibitemStop [0]{}%
\providecommand \bibitemNoStop [0]{.\EOS\space}%
\providecommand \EOS [0]{\spacefactor3000\relax}%
\providecommand \BibitemShut  [1]{\csname bibitem#1\endcsname}%
\let\auto@bib@innerbib\@empty
\end{thebibliography}%


\begin{thebibliography}{0}

\bibitem{Ashcroft76}
N.W. Ashcroft and N.D. Mermin, {\it Solid State Physics}, (Saunders College, 1976).

\bibitem{Imada98}
M. Imada, A. Fujimori, Y. Tokura, Rev. Mod. Phys. {\bf 70}, 1039 (1998).

\bibitem{Anderson58}
P. Anderson, Phys. Rev. {\bf 109}, 1492 (1958).

\bibitem{Kondov11}
S. S. Kondov, W. R. McGehee, J.J. Zirbel, and B. DeMarco,  Science {\bf 334}, 66 (2011); F. Jendrzejewski, A. Bernard, K. M\"uller, P. Cheinet, V. Josse, M. Piraud, L. Pezz\'e, L. Sanchez-Palencia, A. Aspect, and P. Bouyer, arXiv:1108.0137v2 (2011); with a mapping to a quasiperiodic kicked rotor: J. Chab\'e, G. Lemari\'e, B. Gr\'emaud, D. Delande, P. Szriftgiser, and J.C. Garreau, Phys. Rev. Lett. {\bf 101}, 255702 (2008).



\bibitem{Kravchenko04}
S. V. Kravchenko and M. P. Sarachik, Rep. Prog. Phys. {\bf 67}, 1 (2004).

\bibitem{Goldman98}
A.M. Goldman and N. Markovi\'c, Phys. Today {\bf 51}, 39 (1998).

\bibitem{Pan01}
S. H. Pan, J. P. O'Neal, R. L. Badzey, C. Chamon, H. Ding, J. R. Engelbrecht, Z. Wang, H. Eisaki, S. Uchida, A. K. Gupta, K.-W. Ng, E. W. Hudson, K. M. Lang, and J. C. Davis, Nature {\bf 413}, 282 (2001).

\bibitem{Fisher90}
M.P.A. Fisher, G. Grinstein,  S.M. Girvin, Phys. Rev. Lett. {\bf 64}, 587 (1990).

\bibitem{Bollinger11}
A. T. Bollinger, G. Dubuis, J. Yoon, D. Pavuna, J. Misewich, I. Bozovi\'c, Nature {\bf 472}, 458 (2011).

\bibitem{Krauth91}
W. Krauth, N. Trivedi, and D. Ceperley, Phys. Rev. Lett. {\bf 67}, 2307 (1991); F. Lin, E.S. S{\o}rensen, and D.M. Ceperley, Phys. Rev. B {\bf 84}, 094507 (2011); S. G. S\"oyler, M. Kiselev, N. V. Prokof'ev, and B. V. Svistunov,  Phys. Rev. Lett. {\bf 107}, 185301 (2011).


\bibitem{Fallani07}
L. Fallani, J.E. Lye, V. Guarrera, C. Fort, M. Inguscio, Phys. Rev. Lett. {\bf 98}, 130404 (2007); B. Deissler,  M. Zaccanti,  G. Roati,  C. DErrico,  M. Fattori,  M. Modugno,  G. Modugno, and  M. Inguscio, Nature Physics {\bf 6}, 354 (2010).

\bibitem{Chen08}
Y.P. Chen, J. Hitchcock, D. Dries, M. Junker, C. Welford, and R.G. Hulet, Phys. Rev. A, {\bf 77}, 033632 (2008); D. Cl\'ement, P. Bouyer, A. Aspect, and L. Sanchez-Palencia, Phys. Rev. A {\bf 77}, 033631 (2008). 


\bibitem{Gadway11}
B. Gadway, D. Pertot, J.Reeves, M. Vogt, and D. Schneble, arXiv:1107.2428v2 (2011).

\bibitem{White09}
M. White, M. Pasienski, D. McKay, S.Q. Zhou, D. Ceperley, and B. DeMarco, 
Phys. Rev. Lett. {\bf 102}, 055301 (2009); M. Pasienski, D. McKay, M. White and B. DeMarco, Nature Physics {\bf 6}, 677 (2010).

\bibitem{Berezinskii72}
V. L. Berezinskii, Sov. Phys. JETP {\bf 34}, 610 (1972); J. M. Kosterlitz and D. J. Thouless, J. Phys. C {\bf 6}, 1181 (1973); D. J. Bishop and J. D. Reppy, Phys. Rev. Lett. {\bf 40}, 1727 (1978).

\bibitem{Hadzibabic06}
Z. Hadzibabic, P. Kr\"uger, M. Cheneau, B. Battelier, and J.
Dalibard, Nature {\bf 441}, 1118 (2006); Z. Hadzibabic and J. Dalibard, Riv. Nuo. Cim., {\bf 34}, 389-434 (2011).

\bibitem{superfluid}
Actually, the superfluid fraction has never been directly measured in 2D quantum gases. 

\bibitem{Pilati09}
S. Pilati, S. Giorgini, and N. Prokof'ev, Phys. Rev. Lett.  {\bf 102}, 150402 (2009);
S. Pilati, S. Giorgini, M. Modugno and N. Prokof'ev, New J. Phys.  {\bf 12}, 073003 (2010).

\bibitem{Abrahams79} E. Abrahams, P.W. Anderson, D.C.
Licciardello, and T.V. Ramakrishnan, Phys. Rev. Lett. {\bf 42},
673 (1979).

\bibitem{Blatter94}
G. Blatter, M. Y. Feigel'man, V. B.Geshkenbein, A. I. Larkin, and V. M. Vinokur, Rev. Mod. Phys. {\bf 66}, 1125 (1994).


\bibitem{Plisson11}
T. Plisson, B. Allard, M. Holzmann, G. Salomon, A. Aspect, P. Bouyer, and T. Bourdel, Phys. Rev. A {\bf 84}, 061606(R) (2011).

\bibitem{Gerbier03}
F.Gerbier, J.H. Thywissen, S. Richard, M. Hugbart, P. Bouyer and A. Aspect, Phys. Rev. A {\bf 67}, 051602 (2003); S. Richard, F.Gerbier, J.H. Thywissen, M. Hugbart, P. Bouyer and A. Aspect, Phys. Rev. Lett. {\bf 91}, 010405 (2003), and references therein.


\bibitem{Clement06}
D. Cl\'ement, A.F. Var\'on, J.A. Retter, L. Sanchez-Palencia, A. Aspect and P. Bouyer, New J. Phys. {\bf 8}, 165 (2006).

\bibitem{Robert10}
M. Robert-de-Saint-Vincent, J.-P. Brantut, B. Allard, T. Plisson, L. Pezz\'e, L. Sanchez-Palencia, A. Aspect, T. Bourdel, and P. Bouyer, Phys. Rev. Lett. {\bf 104}, 220602 (2010).

\bibitem{Prokofev01}
N. Prokof'ev, O. Ruebenacker, and B. Svistunov, Phys. Rev.
Lett. {\bf. 87}, 270402 (2001); N.~Prokof'ev and B.~Svistunov, Phys. Rev. A
{\bf 66}, 043608 (2002).

\bibitem{Rath10}
S. P. Rath, T. Yefsah, K.J. G\"unter, M. Cheneau, R. Desbuquois, M. Holzmann, W. Krauth, and J. Dalibard, Phys. Rev. A {\bf 82}, 013609 (2010).

\bibitem{Holzmann08}
M. Holzmann and W. Krauth, Phys. Rev. Lett. {\bf 100}, 190402 (2008).

\bibitem{Yefsah11}
T. Yefsah, R. Desbuquois, L. Chomaz, K.J. G\"unter, and J. Dalibard, Phys. Rev. Lett. {\bf 107}, 130401 (2011).

\bibitem{correspondence}
Experimentally, the correspondence between $N_0/N$ and the width of the momentum distribution $n(k)$ is only weakly modified in the presence of disorder. This correspondence can be found in \cite{Plisson11}.

\bibitem{Aleiner10}
I. L. Aleiner, B. L. Altshuler, and G. V. Shlyapnikov, Nature Physics {\bf 6}, 900 (2010).

\bibitem{Weinrib82}
L. N. Smith and C. J. Lobb, Phys. Rev. B {\bf 20}, 3653 (1979); A. Weinrib, Phys. Rev. B. {\bf 26}, 1352 (1982).

\bibitem{estimate}
We are not aware of any calculations of the critical parameters of a correlated disordered system at the quantum phase transition in a continuous 2D system. A rough estimate may be obtained from disordered 2D Bose-Hubbard models. Since the typical microscopic distance of our experimental setup is given by $a \approx (\sigma_x \sigma_y)^{1/2} \approx \lambda_{dB}$, the relevant lattice parameters are $t=\hbar^2/2ma^2$  and $U=\hbar^2 \tilde{g}/ma^2$. The critical on-site disorder strength for the Bose glass transition \cite{Krauth91} $\Delta_c \simeq 20 t \sqrt{U/t}$ then corresponds to $\approx 50$\,nK for our parameters.


\bibitem{Beeler11}
M.C. Beeler, M.E.W. Reed, T. Hong, and S.L. Rolston, arXiv:1111.1316v1 (2011).





\end{thebibliography}
\end{document}